\documentclass[letter, twocolumn]{article}
 
\usepackage{times}
\usepackage{epsfig}
\usepackage{subfigure}
\usepackage{graphics}
\usepackage{graphicx}
\usepackage{multirow}
\usepackage[margin=2.7cm]{geometry} 
\usepackage[latin1]{inputenc}

\newcommand{\institutions}{\mbox{
      \begin{minipage}{7cm}
        \begin{center}
          $^{\dag}$ Computer Science Department\\
          Universidade Federal de Minas Gerais \\
          Belo Horizonte - Brazil \\
          \{lhg, barra, virgilio, jussara\}@dcc.ufmg.br
        \end{center}
      \end{minipage}
      \begin{minipage}{7cm}
        \begin{center}
          $^{\ddag}$ Computer and Computational Sciences\\
          Los Alamos National Laboratory \\
          Los Alamos - USA\\
          lmbett@lanl.gov
        \end{center}
      \end{minipage}}}

\begin{document}

\title{Comparative Graph Theoretical Characterization of Networks of Spam and Legitimate Email}

\author{
  {\bf Luiz H. Gomes$^{\dag}$, Rodrigo B. Almeida$^{\dag}$, Luis
    M. A. Bettencourt$^{\ddag}$,} \\
  {\bf Virgílio  Almeida$^{\dag}$, Jussara M. Almeida$^{\dag}$} 
\\[0.3cm]
\institutions
}

\maketitle

\begin{abstract}
Email is an increasingly important and ubiquitous means of 
communication, both facilitating contact between private individuals and
enabling rises in the productivity of organizations.  However the
relentless rise of automatic unauthorized emails, a.k.a. spam is eroding away
much of the attractiveness of email communication. Most of the attention
dedicated to date to spam detection has focused on the content of the emails
or on the addresses or domains associated with spam senders. Although methods
based on these - easily changeable - identifiers work reasonably well they miss
on the fundamental nature of spam as an opportunistic relationship,  very
different from the normal mutual relations between senders and recipients of
legitimate email. 
%In this sense spam traffic is fundamentally asymmetric, and as
%such distinct from regular email.  
Here we present a comprehensive graph
theoretical analysis of email traffic that captures these properties quantitatively. 
We identify several simple metrics that serve both to distinguish between 
spam and legitimate email and to provide a statistical basis for models of spam traffic.    
\end{abstract}

\section{Introduction}
\label{intro}

Spam  is quickly becoming the leading threat to  the viability of email 
as a means of communication and a leading source of fraud and other criminal
activity worldwide. Much is known about spam traffic. According to the Spamhaus
project~\cite{spamhaus} the vast majority of spam emails presently originate
in the USA and China, hosted by well known ISPs and generated  by identified
individuals. Nevertheless an increased effort in criminal investigation and
waves of high profile legislation have not yet succeeded at reducing the
relentless increase in spam traffic~\cite{NYTarticle}, which now accounts for
about $83\%$ of all incoming emails, up from $24\%$ in January 2003
\cite{messageLabs}.   

It is often said that the problem of spam email is that it is an extremely
asymmetric threat. While it is technically easy and very cheap to send a spam
email it requires sophisticated organization and much higher costs at the
receiving end to sort out legitimate emails from junk. 

This asymmetry is of course not directly manifest in the sender's email
address, on the domain he/she uses, nor certainly on the simplest
characteristics of the message (e.g. its size). It is rather a property of
structural relationships - spammers tend to be senders to a socially unrelated
set of receivers - while legitimate email tends, instead, to reflect the variety of
mutual personal, professional, institutional ties among people. Thus by
identifying the comparative structural and dynamical nature of email traffic,
we expect to find good discriminators between normal email and spam
traffic. The goal of this work is to present the modeling of email - legitimate
and spam - traffic as networks,  in order to identify graph theoretical metrics
that can be used to differentiate between the two. 
%spam and non-spam. 
We are also interested in providing a unified view of several metrics characterizing
the relationships between senders/recipients and of their evolution for legitimate
and spam traffics in order to formulate, in the future, a predictive model of spam
dissemination.  

Our study goes beyond several recent analyses~\cite{emailnetcombat,spammachines} 
on the graphical nature of spam traffic. We deal with a different database, involving a much larger number of users and messages, and analyze a wider set of metrics, both static and
dynamic. We will show that there is no single graphical  metric that unequivocally distinguishes
 between legitimate and spam email. There
are, however, several graph theoretical measures that can be combined into a
probabilistic spam detection framework. 
%We will also be able to rank graphical measures according to their
%ability to differentiate spam from non-spam. 
These are then identified as candidates for the construction of a future spam
filtering algorithm.   

The remaining of this paper is organized as follows. In section~\ref{graphs} we
introduce the modeling of email traffic in terms of two graph classes and
present the types of metrics to be studied. Section~\ref{sec:workloads} gives
several global properties of our workload. We evaluate the several metrics, for
each of the two graph classes, in Section~\ref{sec:networks-spam-vs}. In
Section~\ref{sec:related-work} we present related work. Finally, we present our
conclusions in section~\ref{sec:concl-future-works} and discuss open questions
left for future work.  

\section{Graph-Based Modeling of E-mail Workloads}
\label{graphs}

In order to characterize spam email traffic versus  non-spam we define two
types of graphs: a {\it user graph} and a {\it domain graph}. 
%These graphs are built based on information derived from the aggregated traffic several users of
%the same domain have received/sent. 
The vertices of the {\it user graph} are email senders and recipients
present in our log. An email sent  by A to receiver B results in a link
between A and B. The {\it domain graph} has as vertices the domains of the
external senders to the local domain being analyzed, and users if inside the
local domain. Its construction is similar to the {\it user graph} but sets
of users external to the local domain who share an external domain are
aggregated together into a single node. Note also that the domain graph is
a simpler bipartite graph and not all characteristics studied will be present in it.

The edges of both graphs can take one of four forms: directed or undirected; 
binary\footnote{If {\em any} message was sent from A to B,
  over the observation time a link is established.} (or unweighted) or weighted (e.g. by the
number of emails exchanged or by the total size of the emails exchanged in
bytes).  These options cover most of the possibilities for direct
graphical construction out of the email logs at our disposal (described in
Section~\ref{sec:workloads}).  

The user graph is in principle the most useful in identifying the 
individual nature of users as spam or non-spam senders. In some cases these 
characteristics extend to the whole external domain (particularly if the
spammer changes his name\footnote{The first part of the address, located before
the @.} more often than its domain) and the domain graph produces a
useful aggregation of the user data. We believe that user graphs will be more
effective in identifying senders of non-spam since spam senders tend to change their
full email address very frequently. 
% Luis: penso que isso nao e necessario: 
%The reason for only considering
%external domains in the domain graph comes from the fact that the consideration
%of the local domain would create a graph that is centralized in one domain
%instead of represent the relations distributed through several domains.

%Since spam messages generally contain
%commercial solicitations we hope to distinguish between different
%business entities more readily through the domain graph. 

The user or domain graphs can be constructed exclusively out of spam
traffic, non-spam traffic, or the aggregate set of all emails. Some 
of the graph theoretical properties studied below will be analyzed in terms of 
the graphs formed when considering the different traffics separatelly while
others will be evaluated on selected nodes from the aggregated traffic. 
The selected nodes represent senders in the aggregated graph and can be divided
in two classes - spam and non-spam -  based on  the type of emails they
send. These classes do not form disjoint sets, see Table~\ref{tab_extint}. Since
we are analyzing nodes in terms of the email types 
they send, we will not present an analysis of the edges (traffic)
comming in. In other words we will not attempt to identify spammers from the
set of emails that are sent to them, simply because the statistical properties of such messages 
are clearly less significant as those of the messages they send out.

% one following
%the types of messages they send and the other considering whether they are
%internal or external addresses. Considering the messages they send, nodes can
%be divided in the following classes: {\it spam nodes} are the nodes that have
%sent at least one spam message, {\it non-spam} are the group that have sent at
%least one non-spam message and the third and last class is formed by the
%intersection of the previous one. 

Given these two graph constructions we will analyze two types of properties: 
(i) structural and (ii) dynamical. The former capture the structure of
social relationships between users exchanging emails, while the latter
relate to how graphical  properties evolve over time. As we shall show below there 
are distinct independent signatures of spam traffic in both structural and dynamical 
properties. As a consequence they should be taken together to generate a better 
detection procedure. 

\section{E-mail Workloads}
\label{sec:workloads}

\begin{table}[th!]
\centering
\footnotesize
\label{tab:bst}
\begin{tabular}{|l|r|r|r|} \hline
{\bf Measure}                           & {\bf Non-Spam}& {\bf Spam}         &  {\bf Aggregate}                 \\
\hline
\# e-mails         &  336,580        & 278,522        &    615,102                        \\ \hline
Size of e-mails       &  11.00 GB        & 1.70 GB        &   12.71 GB                        \\ \hline
\# sender users    & 94,985        &    170,664     &  263,144        \\ \hline
\# sender domains  & 20,414       &  48,087      &  59,971           \\ \hline
\# recipients      &  26,450   &  12,867 &  35,471 \\ \hline
\end{tabular}
\caption{Workload summary}
\label{tab_summary}
%%%%\vspace{-0.15cm}
\end{table}

The construction of the graphs introduced in Section~\ref{graphs} is subject
to several practical constraints. Our knowledge of email traffic comes from 
Postfix logs of the central SMTP incoming/outgoing servers of an academic
department from a large University in Brazil. Incoming emails only contain
the recipients internal to the department's  domain. 
Outgoing emails contain the full list of recipients. Moreover our data set does 
not contain information about emails exchanged between users external to the domain. 
% but has information between the
%exchanged messages between external users and internal ones.  

The logs were collected between 11/18/2004 and 12/31/2004 and contain the following
data for each email: (i) received time and date; (ii) a reject flag,
indicating whether connection was rejected during e-mail acceptance (iii) Size
of email\footnote{Only for the accepted emails.}; (iv) sender address; (v)
list of recipients and (vi) a spam flag, indicating if it was classified as
spam or not by Spam-Assassin~\cite{spamassassin}. The logs were sanitized and
anonymized to protect the users' privacy. Statistical characteristics of the
workload are in agreement with previous email traffic
analyses~\cite{gomes,spam,priority}. Table~\ref{tab_summary} 
summarizes the data set.
%Data were collected from
%11/18/2004 to 
%12/31/2004, 

Spam-Assassin~\cite{spamassassin} is a popular spam filtering software that
detects spam emails  based on a changing set of user-defined rules. These
rules assign scores to each received e-mail based on the presence in the 
subject or in the e-mail body of one or more pre-categorized
keywords. Spam-Assassin also uses other rules based on email size 
and encoding.
%, which categorize messages
%larger than a pre-defined size as legitimate non-spam e-mails.
Highly ranked emails, according to these criteria, are flagged as spam. 
%Spam-Assassin also uses size-based rules,
%which categorize messages larger than a pre-defined size as legitimate non-spam e-mails. 

\section{Spam Networks vs. Legitimate Email Networks}
%\section{Networks of Spam vs. Regular Email}
\label{sec:networks-spam-vs}

\begin{table}[th!]
\footnotesize
\centering
\begin{tabular}{|l|rr|rr|rr|} \hline
{\bf Type} & \multicolumn{2}{|c}{\bf External}& \multicolumn{2}{|c|}{\bf Internal}  \\ \hline
Spam       &  169931  & (277535)          & 733  & (987)        \\ \hline
Non-Spam   &  93666  &(186607)        & 1319   & (186607)   \\ \hline
Spam \&    & \multirow{2}{*}{2366}   & \multirow{2}{*}{(-)} & \multirow{2}{*}{139} & \multirow{2}{*}{(-)} \\ 
Non-Spam       & & & &         \\ \hline
Total      & 263231 & (462142)       & 1913 & (152960)   \\ \hline
\end{tabular}
\caption{Number of unique email addresses by origin (internal or external to the domain) and classified 
as spam, non-spam or both. Numbers in parentheses indicate the total number of
emails sent by each class.} 
\label{tab_extint}
%%%%\vspace{-0.15cm}
\end{table}

Although spam emails originate principally from users outside the local
domain spam senders use several techniques to falsify or steal local addresses
(e.g. crawling the web for email addresses available at web pages,
network sniffing, name dictionaries). As a result spam email does originate
from the local domain both from real users and from forged ones. 
This mixing between regular email users and spam senders can lead to more 
complex email networks than might have been naively expected and poses a
challenging problem for detection. 

Table~\ref{tab_extint} summarizes the number of addresses and emails by node
classes and by internal or external origin. Node classes are as defined in Section~\ref{graphs} plus a third category -Spam \& Non-Spam - which is the intersection of the former two.  
The size of this overlap shows the impact of email address spoofing. 

%Table~\ref{tab_extint} summarizes the number of addresses and messages sent as part of 
%spam, non-spam email traffic, with both internal origin or external. Because there is 
% mixing between spam and non-spam senders this classification specifically means addresses 
%that sent at least one spam message are classified as spam, and those who sent at least 
%one non-spam message as non-spam. The overlap between those two sets are those addresses 
%that sent at least one spam {\em and} one non-spam message.  
%Measuring the number of addresses in this class allows us to assess the impact of 
%email address spoofing.

Most emails originate outside the domain. In our log most outside users are
spam senders and account for the majority of the emails. Because it is 
very easy for a spammer to forge an address
spam senders use many addresses simultaneously and/or frequently switch between
them. This strategy is visible in our database as non-spam internal
users send many more emails per user than spam internal users. We expect that
this is a general feature of spam versus non-spam traffic. %, but our log does not allow us to comment on external senders as we only see a small part of their target receivers.  

The number of spam senders that are internal is very small. The  fraction of
these that send exclusively spam is 81\%. These  addresses correspond presumably to
internal emails that have been forged and do not actually exist\footnote{This suggests that a simple effective way to filter out spam originating from internal domain addresses is to verify that they correspond to an existing user.}. The remaining
addresses send both spam and non-spam and are probably genuine users whose
addresses have been spoofed.   

\subsection{Structural analysis of spam vs. non spam email graphs}

%\subsubsection{Degree}

\begin{figure}[ht!]
  \footnotesize
  \begin{center}
    {\includegraphics[width=150pt]{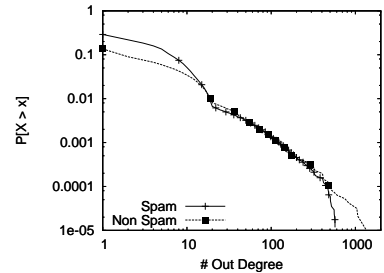}} \\[0.1cm]
    (a) User Graph \\[0.2cm]
    {\includegraphics[width=150pt]{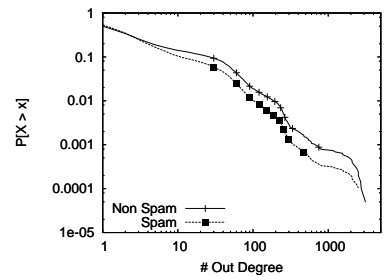}} \\[0.1cm]
    (b) Domain Graph
  \end{center}
  \caption{Distribution of the node degrees for sender classes in the
    aggregated graphs. }
  \label{fig:degree-ccdf}

%%%%\vspace{-0.15cm}
\end{figure}

One of the most common structural measures analyzed in complex networks
is the distribution of the number of the incoming and outgoing node
connections, or degree~\cite{socialxother,emailnetwork,technologicalnetworks}. 
Figure~\ref{fig:degree-ccdf} shows the distribution of the out-degrees
of the different sender classes for the user and domain graphs. 

The out degree distributions approximately follow a power law ($C/x^{\alpha}$).
By using a simple statistical linear regression we estimated the 
exponent $\alpha$ that best models the data. For the user graph we obtained $\alpha$ = 1.497 (with $R^2$ = 0.965.) for spam senders
and $\alpha$ = 1.359 ($R^2$ = 0.981) for non-spam senders. 
We conclude that the spam sender's out degree distribution is slightly more skewed. 
We conjecture that this is because spammers have a limited knowledge of the set
of users in each specific domain. Since in our analysis we only observe a fraction
of the spammers' lists (the one composed by the messages sent to the domain
studied) there are no spammers with recipients' lists as large as those found 
for non-spam senders.
%there are not spammers with list as large as the biggest ones found in non-spam traffic.
% Moreover,
% the distribution for the out degree of non spam is better modeled by a power
% law.  

Degrees from 1 to near 20 are much more probable for spam senders than for non spammmers, while very large degrees are more likely in non-spam. There is no difference between the two
sender classes in the body of the distribution, for degrees from about 20 to
400. The mean out-degrees, are 3.56 and 1.63 for non-spam and spam,
respectively (see  Table~\ref{tab_extint}).  

In the domain graph the out-degree distribution shows a much higher
probability for nodes with low out-degree in spam traffic than in 
non-spam.

%The in-degree distribution, as shown in Figure~\ref{fig:degree-ccdf}, shows
%that in both graph types nodes in the non-spam class have much  larger
%probability to have higher in-degree. These results confirm the insight that
%spam senders are almost never contacted back by their recipients. We will see other signatures 
%of this effect below. 

%However, in disagreement with the previous observation, we also find that for in-degree near and over 1000, spammers have higher in-degrees. Further analysis of the data
%showed that these addresses are the group of forged internal spam senders that are both senders and as recipients of spam and non spam messages. These addresses have probably been stolen as a  result from dictionary attacks. Thus, although there are not many such addresses their share of the traffic can still be very high.

% comunication reciprocity

\begin{figure}[ht]
\footnotesize
  \begin{center}
    {\includegraphics[width=150pt]{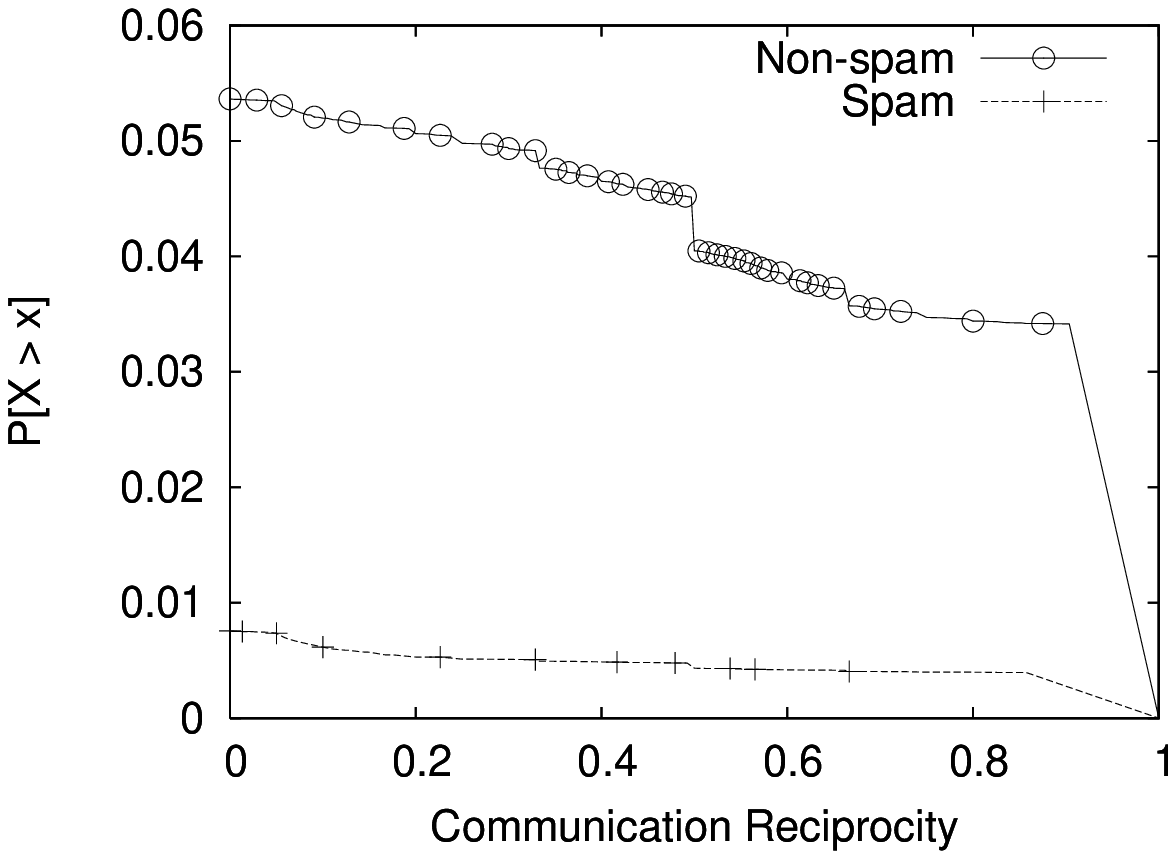}} \\[0.1cm]
    (a) User Graph \\[0.2cm]
    {\includegraphics[width=150pt]{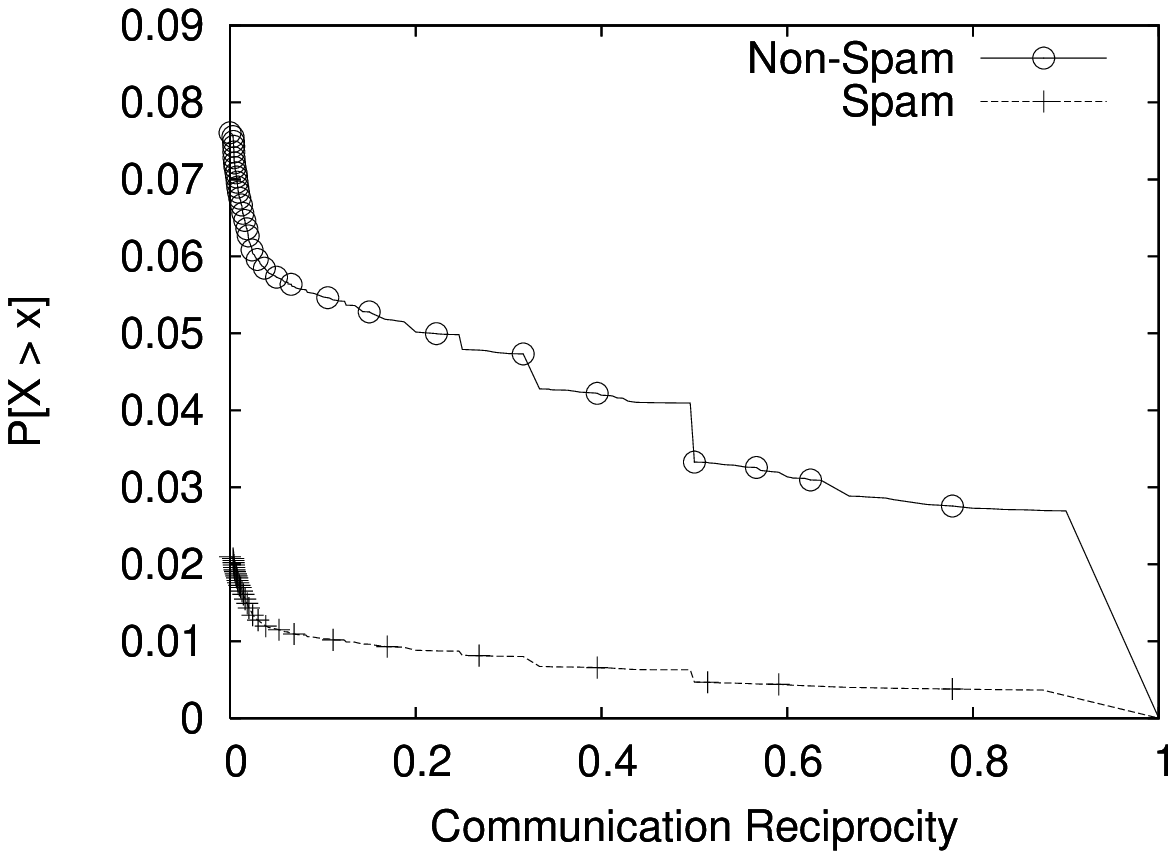}} \\[0.1cm]
    (b) Domain Graph \\[0.2cm]
  \end{center}
  \caption{Distribution of Communication Reciprocity}
  \label{fig:reciprocity}
  
%%%%\vspace{-0.15cm}
\end{figure}

%Moreover we were also interested the asymmetry of the communication existent in
%the way spammer behave with their audience considering the other side of the story,
%i.e. if we could find spammers by their signature in terms of the destinations
%of their messages and their replies.
 
%The asymmetry between in and out traffic that is observed in the degree distribution of spammmers and non-spammers is best captured by a few direct metrics. 
In order to evaluate discrepancies between in and out sets of addresses for a
given node we create a simple metric called Communication Reciprocity (CR) of $x$ as: 
\begin{equation}
  CR(x) = \frac{|OS(x) \cap IS(x) |}{|OS(x)|},
\end{equation}
where $OS(x)$ is the set of nodes that receive a message from node $x$ and
$IS(x)$ is the set of addresses that send messages to $x$.  With our choice of
normalization this metric measures  the probability of a node receiving a
response from each one of his addressees.
%one of its outgoing messages. 

Figure~\ref{fig:reciprocity} shows the distribution of the Communication 
Reciprocity. This metric is able to effectively differentiate users associated
with spam from non-spam.  The grouping of users in the domain
graph makes this differentiation more difficult. However, even in the
domain graph the difference is very clear.

% Barra e LHG: Oi Luis, acho que nao existe essa distincao tao clara assim nao,
% o que acontece eh que as probabilidades para non-spam sao maiores e nao os
% valores propriamente ditos por isso threshold nao vale. 
% {\bf please check this } virtually
%all users associated with spam have a CR, either in the user or the domain
%graph,  below  all  non-spam users. Thus, given enough data,  a simple
%threshold decision suffices to determine if a user is associated either with
%spam or with legitimate traffic. 

%As one would expect, email
%spoofing causes the appearance of a third behavior intermediate between the
%characteristics from spam and non-spam classes. The effect of
%spoofing in the user graph is much more noticeable than  in
%the domain graph.

% CL-DL

The analysis of the communication reciprocity suggests that a strong
signature of spam is its structural imbalance between the set of senders and
receivers associated with a spam sender. However  whenever there is an
imbalance, how many of the unmatched addresses correspond to spam senders? 

\begin{figure}[ht!]
\footnotesize
    \begin{center}
      \begin{tabular}{cc}
      {\includegraphics[width=100pt]{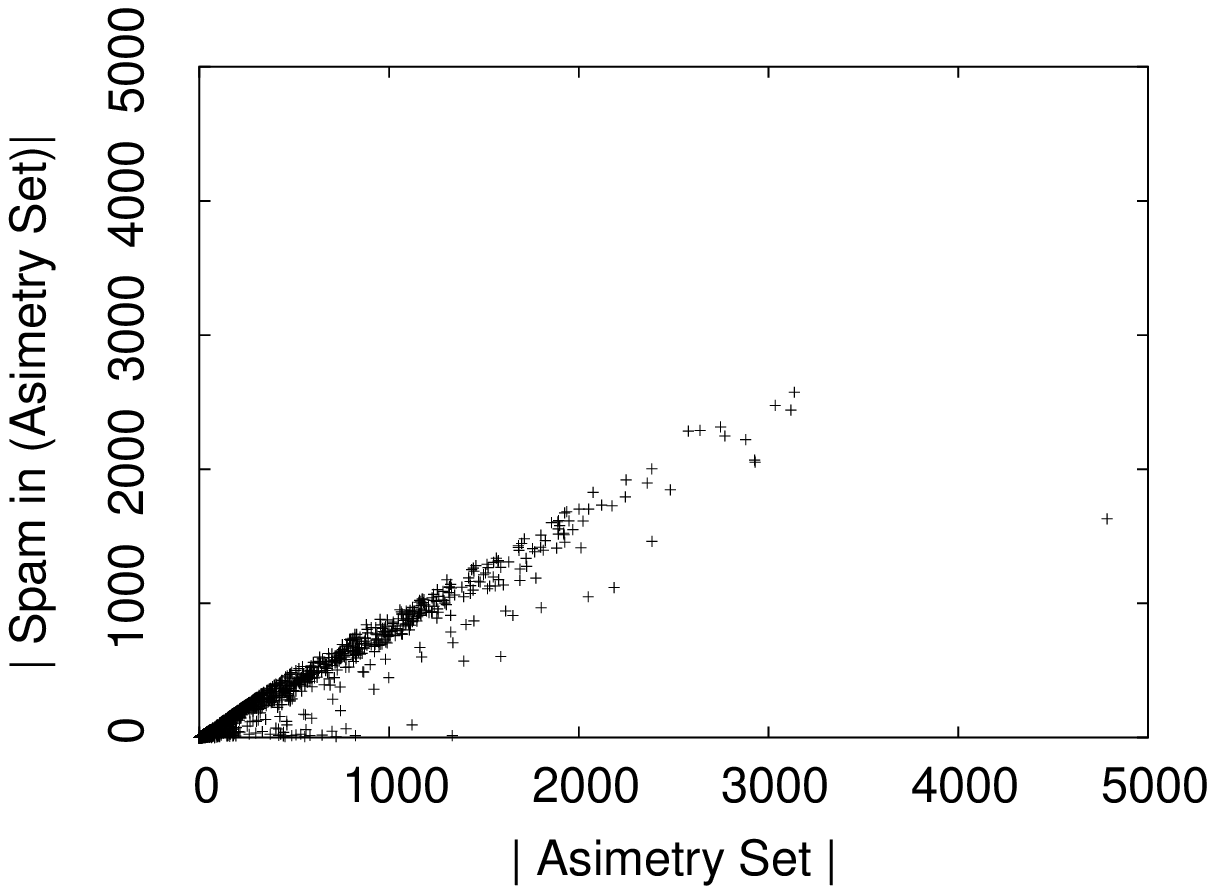}}&
      {\includegraphics[width=100pt]{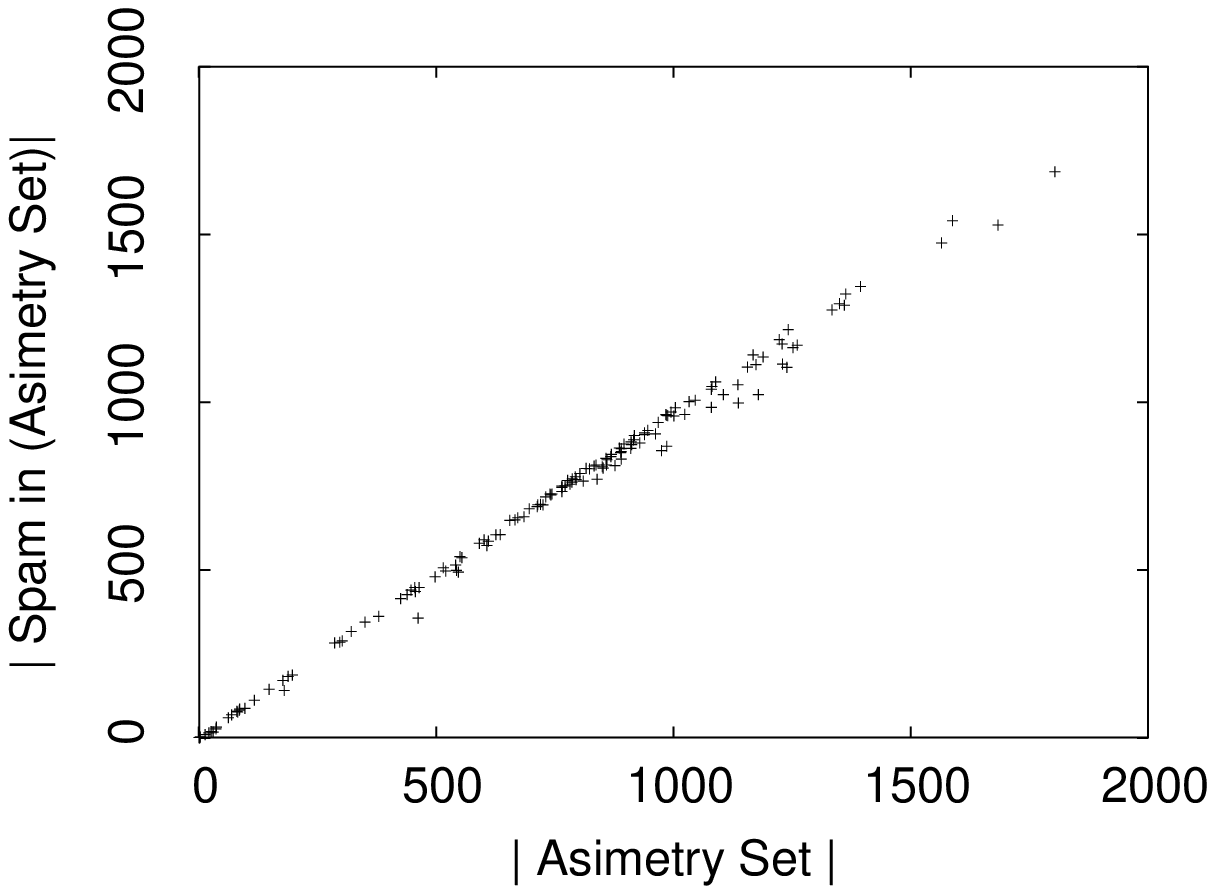}}\\
      {\includegraphics[width=100pt]{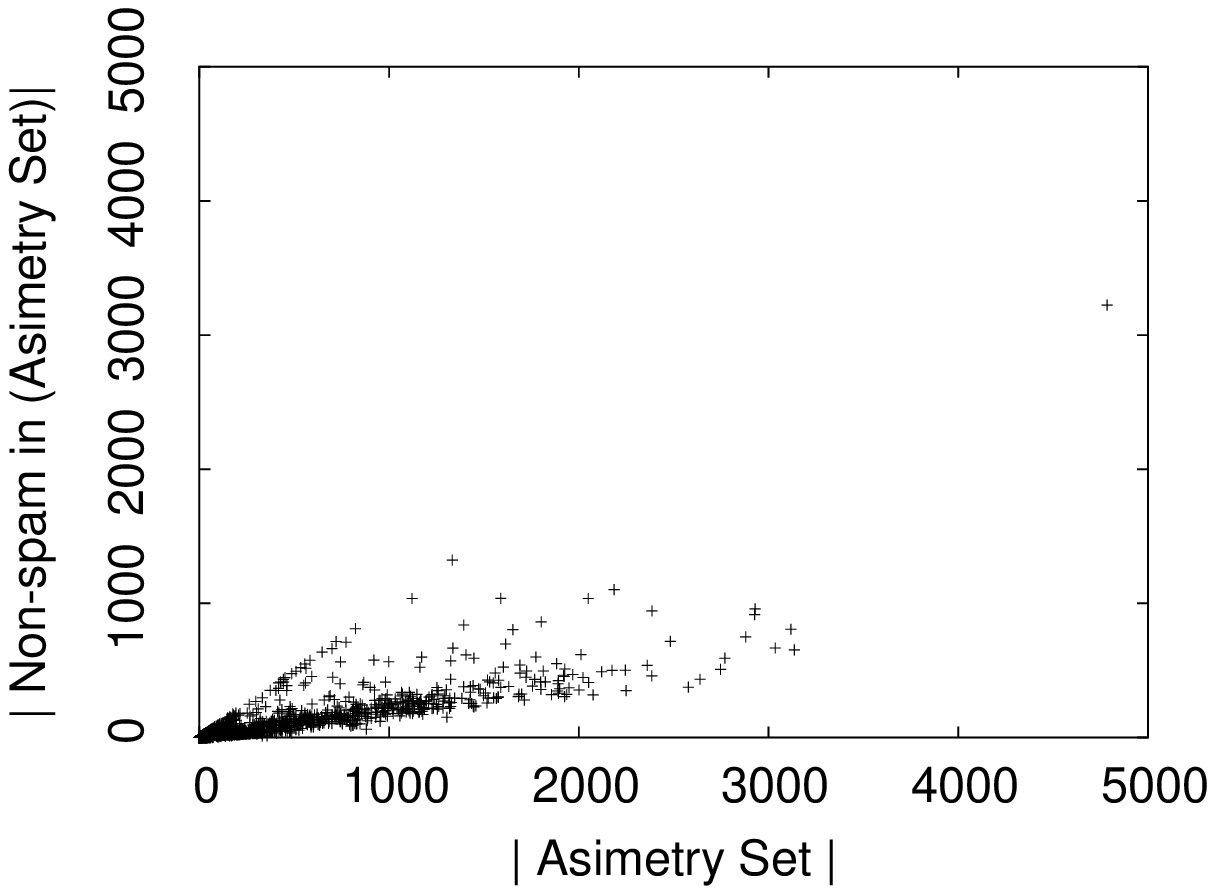}}& 
      {\includegraphics[width=100pt]{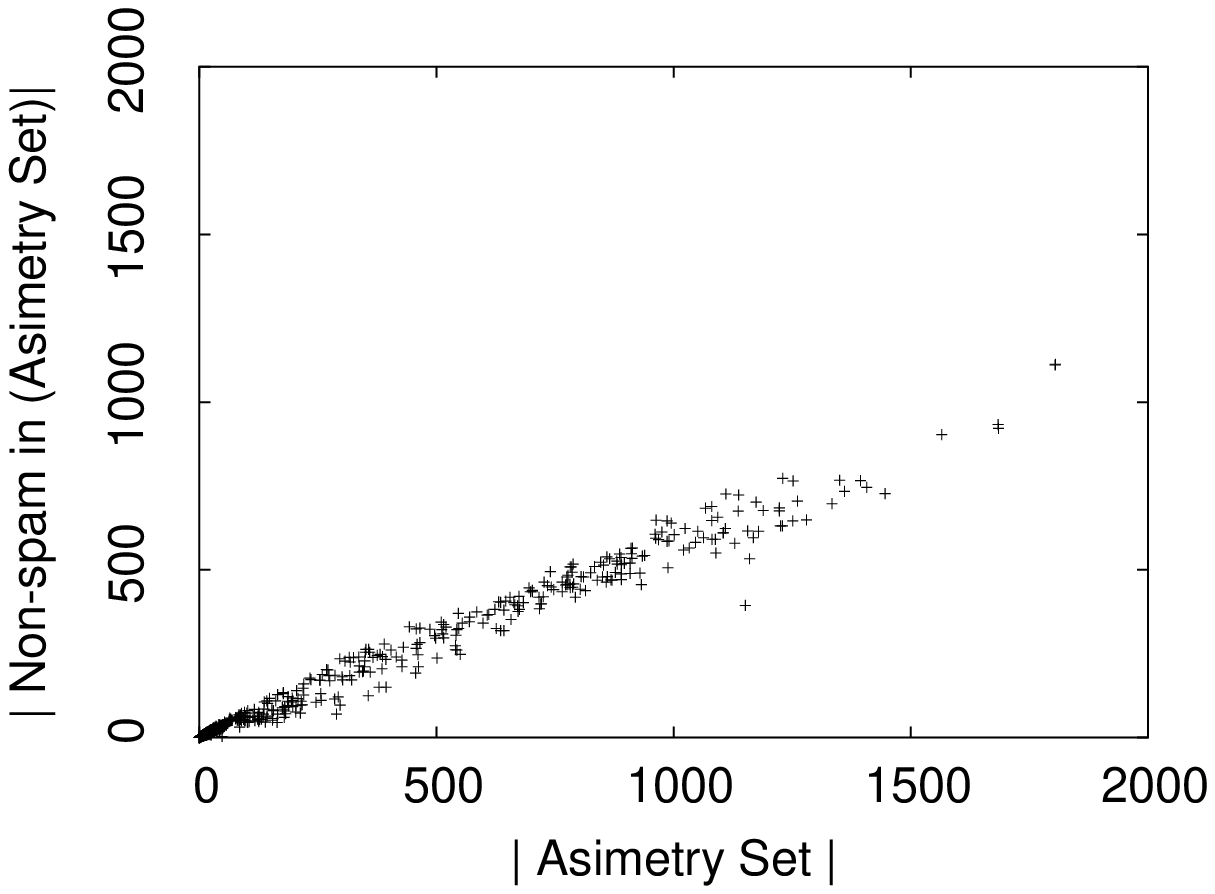}}\\
      (a) User Graph & (b) Domain Graph
      \end{tabular}
%         \end{tabular}
   \end{center}
    \caption{Number of spams/hams in the asymmetry set vs. the number of nodes
      in the asymmetry set}
    \label{fig:corr-cl-dl}

%%%%\vspace{-0.15cm}
\end{figure}

To address this question, let the asymmetry set for a node be the difference of
its in and out sets. Figure~\ref{fig:corr-cl-dl} shows the number of spam
addresses in the asymmetry set versus the size of the asymmetry set itself. The
resulting relation is very well fit by a straight line at $45^o$, showing a
strong correlation between the two numbers. The 
statistical correlation (slope) is $\rho=0.979$ for user graph  and $\rho=0.998$
for the domain graph. So, almost all senders in the asymmetry sets are
spammers indifferently of the graph analyzed. The non spam data is not very well
modeled by a $45^o$ straight line. These correspond to the non spam senders
that were not answered (or to whom we could not see an answer in our log). The correlation is $\rho = 0.8723$ and $\rho = 0.9932$ for the user and domain graphs respectively. As expected from the
result of the spam data the non spam data has a higher correlation for the domain graph. 

 This result can be made sharper if we analyze the correlation between the 
number of spammers in the incoming set of a node and spammers in its asymmetry
set. We find $\rho=0.999$ and $\rho=0.994$ for the user and domain graphs,
respectively. There is a slightly worse correlation in the domain graph. 
We conjecture this is due to the external reliable domains used by spammers  
(e.g. through spoofing and forging techniques). These may not be counted in the
asymmetry set since they are replied through their legitime emails but are
part of the incoming set as spammers.    

These results show that spam messages are almost never replied to, except in cases
of spoofed or forged domains or users' ids and rarely, we assume, intentionally.

Asymmetry sets can in principle be used as a component in a probabilistic spam detection
mechanism. The  arrival of an email from a sender that has already been
contacted by an internal recipient is an indication that it has high
probability of being a non spam email.
%Note that applying this rule for the first time to a new non spam sender will lead to the wrong classification. However, as soon as this message is replied to by an internal sender the symmetry is restored and we stop misclassifying.

% Clustering coefficient

Another common characteristic of social networks is a high average clustering
coefficient (CC)~\cite{evolutionnet}. The CC of a node $n$, denoted $C_n$, is defined as the
probability of any two of its neighbors being neighbors themselves. This metric is 
associated to the number of triangles that contain a node $n$. For an undirected
graph, the maximum number of triangles connecting the $N_n$ neighbors of $n$ is
$N_{n}\times(N_n - 1)/2$.  Thus, the CC measures the ratio between actual
triangles and their maximal value. During clustering coefficient analysis we
only consider the nodes with $N_n > 1$, since this is a necessary condition for
the CC to be nonzero.    

%Table~\ref{tab_cc} shows the average clustering

%\begin{center}
%\begin{table}[th!]
%\centering
%\footnotesize
%\begin{tabular}{|l|r|r|} \hline
%  {\bf Class} & {\bf CC} & {\bf \% of Nodes}\\ \hline
%  Spam/Internal &  0.062  & 53\% \\ \hline
%  Non-spam/Internal & 0.152 & 76\% \\ \hline
%  Spam+Non-Spam/Internal &  0.107 & 98\% \\ \hline
%  Spam/External & 0.053 & 29\% \\ \hline
%  Non-Spam/External & 0.112 & 12\% \\ \hline
%  Spam+Non-Spam/External & 0.110 & 53\% \\ \hline
%\end{tabular}
%\caption{Average clustering coefficient (CC) analysis for the different sender classes in
%  the aggregated undirected graphs.}
%\label{tab_cc}
%\vspace{-0.15cm}
%\end{table}
%\end{center}

\begin{figure}[ht]
\footnotesize
    \begin{center}
      \includegraphics[width=150pt]{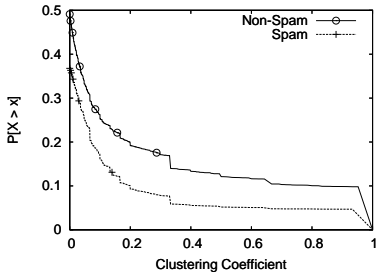}
   \end{center}
    \caption{Distribution of the clustering coefficient for the different
      classes in the aggregated user graph.}
    \label{fig:cc}

%%%%\vspace{-0.15cm}
\end{figure}

%The fact that the social relationships between users that send non-spam messages 
%is much stronger, showing the highest clustering coefficient, was as expected in 
%the analysis of our workload. 

Figure~\ref{fig:cc} shows the distribution for the CC of
nodes in the aggregated graphs. The clustering coefficient measures
cohesion of communication, not only between two users but among {\it friends of
friends}. This is a pervasive characteristic of social relations 
that is absent from spam sender receiver connections. As a result regular email
users have higher CC than spam senders. In terms of the average value, regular
email also has a higher value (0.16 against 0.08).

%cc spammers 0.0831 cv 2.6593
%cc non spammers 0.1563 cv 1.9730

%{\bf Luis. Nao entendemos a definicao acima ``...among users twice
%  removed.''. Isso estah correto? Serah que vc poderia clarificar essa ideia?}

%Moreover, it is important to note the difference between 
%spammers that forge internal emails (Spam+Non-Spam) and the ones
%that send spams from external addresses. The former show an intermediate clustering 
%coefficient, between the higher value displayed by the non-spam group and the 
%lower of the spam group. Clearly this arises as a result of the genuine 
%links that mixed addresses possess, as a result of their fraction of non-spam traffic.
%Thus as a general feature the additional spam traffic that results from having an 
%email spoofed results in a lowering of the clustering coefficient for that node.

%Moreover, the impact of considering edges directionality is
%dependent of the class being analyzed. External nodes have their
%clustering coefficient increased on average when we disregard the
%direction of the edges. Internal nodes, on the other hand, have their clustering coefficient
%decreased. This reflects the fact that we do not have information of messages
%exchanged between external users and increases our believe that it is best to
%evaluate the clustering coefficient as a way to differentiate spammers from
%non-spammers in the directed version of the graph.

% Strongly connected component

Some recent studies~\cite{emailnetcombat} have studied
graphical metrics of the strongly connected components (SCC) of email
graphs. A SCC is a subset of the nodes of a graph, such that one node
can be reached from any other node in the set following edges between them. A
complementary measure to the CC and SCC is the average path length between two
nodes. The CC and average path length properties are generally related to the
so-called small world networks, which display high CC (higher than 
a random graph with the same connectivity) and short path length, usually comparable to $\log N$, where N is the number of nodes in the graph.

In our experiments both the SCC and  the average path
length have not been able to convincingly differentiate spam from legitimate
traffic. 
%{\bf I think this is in part because we neglected the directionality of the links in these 
%diagnostics - didn't we? Maybe we should say that ... That would help the reader get an intuition 
%of why these measures don't work so well, whereas now we really don't have a good reason.}
%Barra e LHG: No trabalho do sujeito ele tbm nao considera a direcionalidade do
%grafo. Note que a implementacao de SCC para grafos direcionados em bem mais
%complexa do que a outra e nao cremos que iria mudar muita coisa
%comparativamente com o trabalho do outro autor dado que ele tbm nao considera
%a direcao.
All of the graphs studied are small world networks to some extent. Also all of
the graphs have giant connected components. Other 
studies have used the clustering coefficient of SCCs to identify spam in
networks  constructed from  the correspondence of a single
user~\cite{emailnetcombat}. However for data from servers that aggregate the
communication between different senders and receivers we find that these metrics do not
suffice to perform a clear identification of spam.  

%% page rank
%
Another interesting structural characteristic of graphs is the probability of visiting a node 
during a random walk through the graphs~\cite{pagerank}. At each step of the random walk we 
need to select the next node to be visited. This can be done in two ways. 
The next node can be randomly selected from the out set of the current node or we can perform a 
jump. For a jump, one of the nodes of the graph is selected randomly as the next node. Note that, this measure is related to node betweenness\footnote{The number of shortest paths that pass throught a node.} since higher node betweenness tends to generate
a higher probability of visitation. Nevertheless this probability is much 
easier to compute than node betweenness for large graphs.  
The probability $P(x)$ of finding a node $x$ in a random walk is computed iteratively as follows:
\begin{equation}
  P(x) = \frac{d}{N} + (1-d)*\sum_{z~\in~IS(x)}{\frac{P(z)}{| OS(z) |}}, 
\end{equation}
where $d$ is the probability of performing a jump during a random walk, $N$ is
the number of nodes in the graph. The parameter $d$ is a dumping factor that can be varied. A value usually used in the literature is 0.15~\cite{pagerank}, that is also the value we use in our measurements.  

\begin{figure}[ht!]
\footnotesize
  \begin{center}
    {\includegraphics[width=150pt]{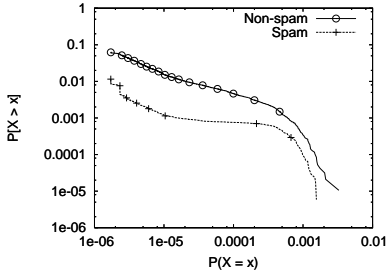}} \\[0.1cm]
    (a) User Graph \\[0.2cm]
    {\includegraphics[width=150pt]{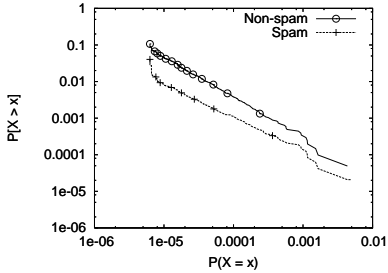}} \\[0.1cm]
    (b) Domain Graph
  \end{center}
  \caption{Distribution of the probability of finding a node during a random walk.}
  \label{fig:pagerank}

%%%%\vspace{-0.15cm}
\end{figure}

The results are shown in Figure~\ref{fig:pagerank}. The difference between spam and non-spam behavior is less
noticeable in the domain graph than in the user graph.  Spam nodes show
generally lower probabilities of being visited, as might have been expected
because of the asymmetry of their communication. Visiting probabilities for
spam nodes in the user graph are localized to the initial and final parts of
the distribution and are less pronounced in the middle range. 

The node visitation probability distributions can be modeled 
by a power law. We estimate the corresponding exponent  at $\alpha= 0.694, \ 1.097$ and $0.975$ for the non-spam component of the user graph, and for the non-spam and spam components in the domain graph, respectively. The $R^2$ associated with the fits varies between 0.959 and 0.998. The $R^2$ for the spam curve of the user graph is 0.853, showing that it is not well modeled by a power law, as visual inspection suggests. 

%\subsection{Dynamical Characteristics}

%\begin{figure}[ht]
%    \begin{center}
%        \begin{tabular}{cc}
%         \subfigure[User Graph]{\includegraphics[width=100pt]{./plots/sndusrrcpusrxnmsg-sna1.obs.ccdf.eps}}&
%         \subfigure[Domain Graph]{\includegraphics[width=100pt]{./plots/snddmnrcpusrxnmsg-sna1.obs.ccdf.eps}}\tabularnewline
%         \end{tabular}
%   \end{center}
%    \caption{Distribution of Flow Between Pairs in Number of Messages}
%    \label{fig:flow_pairs}
%\vspace{-0.15cm}
%\end{figure}

\subsection{Dynamical analysis}
\vspace{0.15cm}

Beyond the structural characteristics of the graphs of spam and non-spam email
other metrics related to the dynamics of communication and graph evolution may
help model spam traffic. 

\begin{figure}[ht!]
\vspace{0.50cm}
\footnotesize  
  \begin{center}    
    \begin{tabular}{cc}      
      \includegraphics[width=100pt]{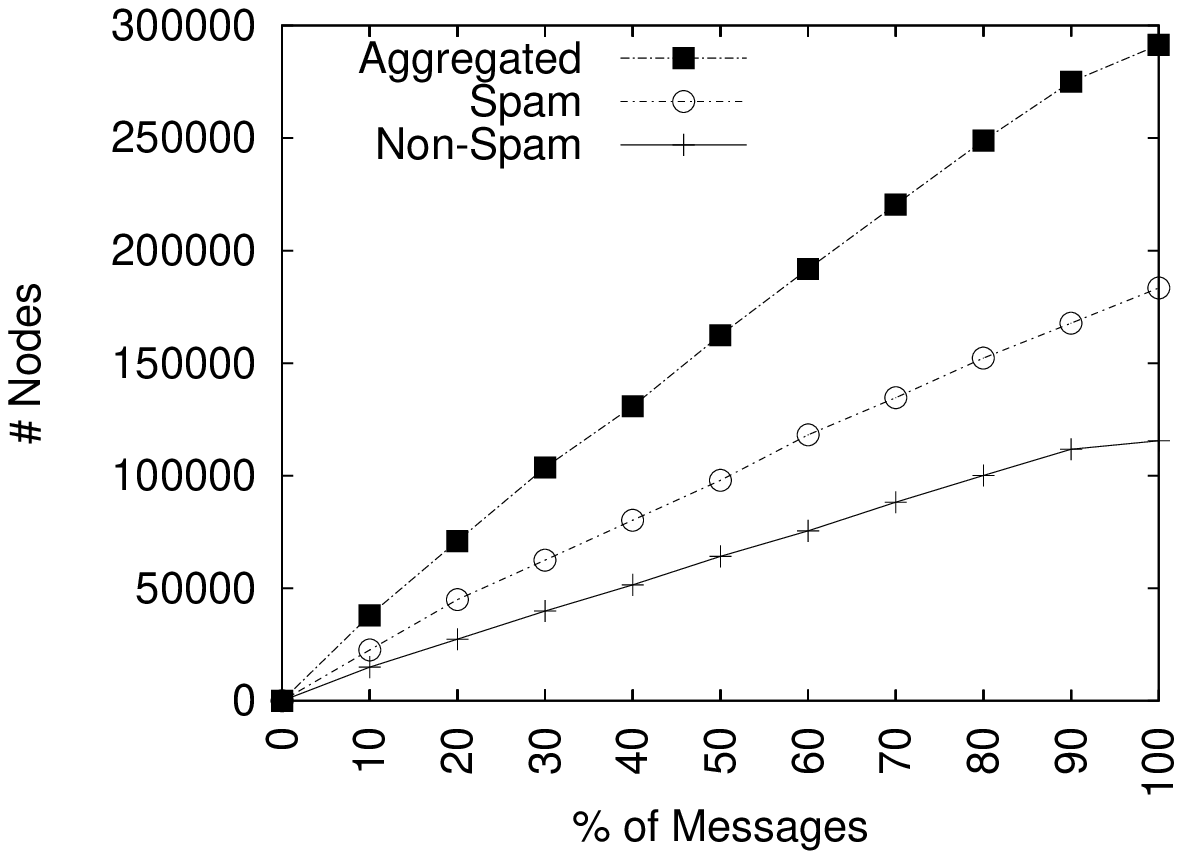}&      
      \includegraphics[width=100pt]{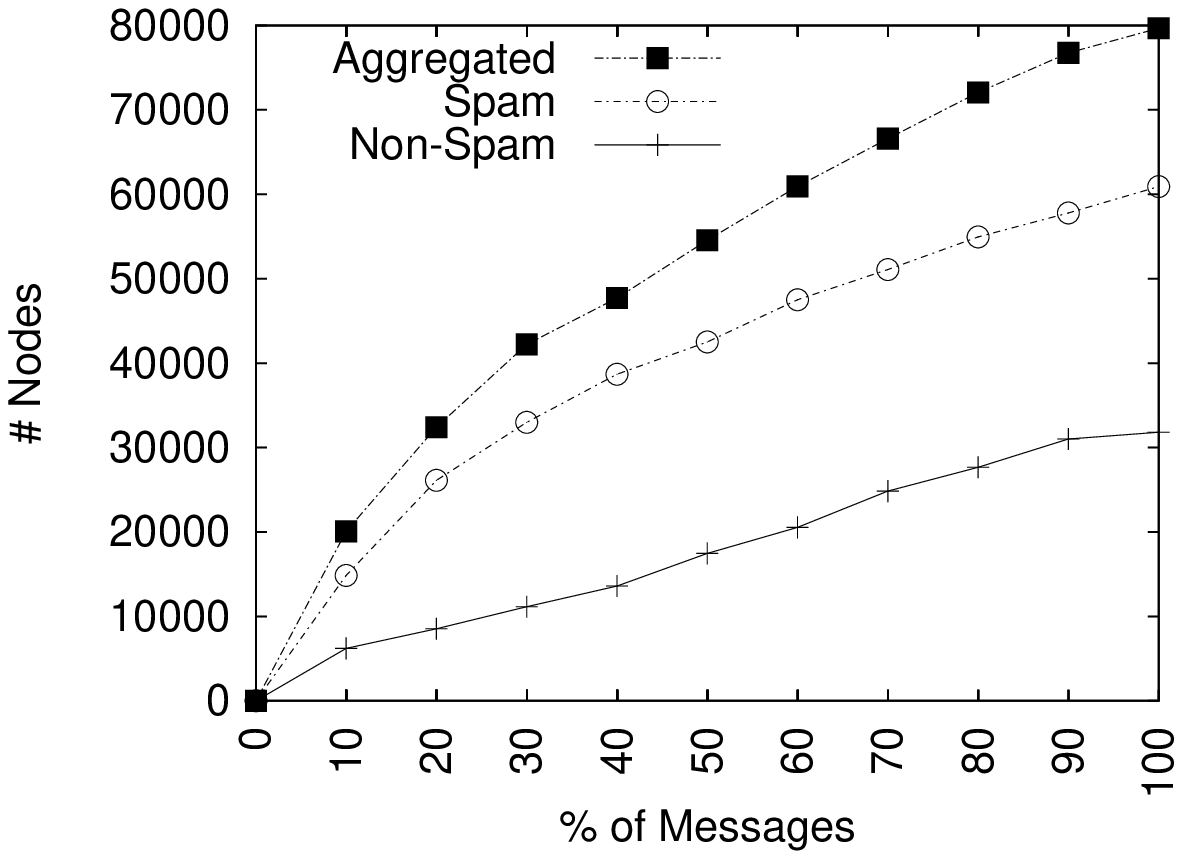} \\      
      \includegraphics[width=100pt]{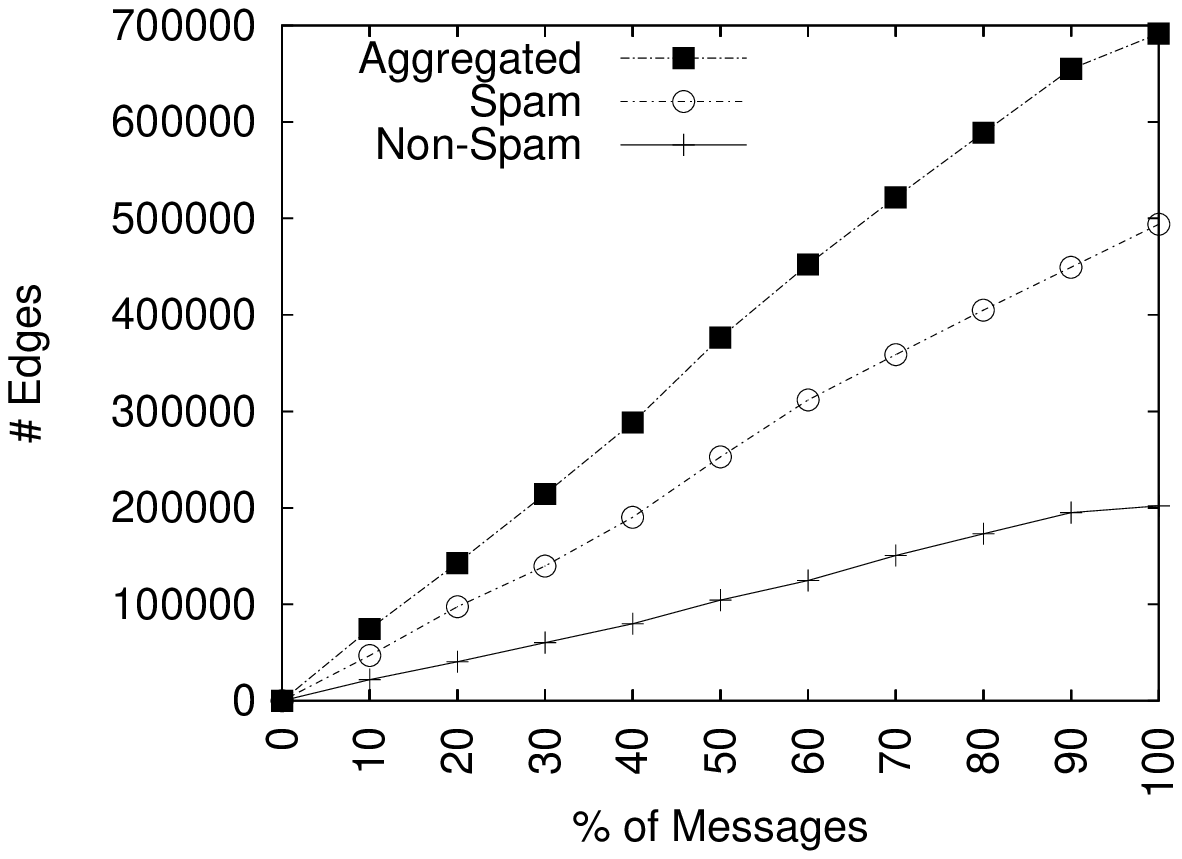}&      
      \includegraphics[width=100pt]{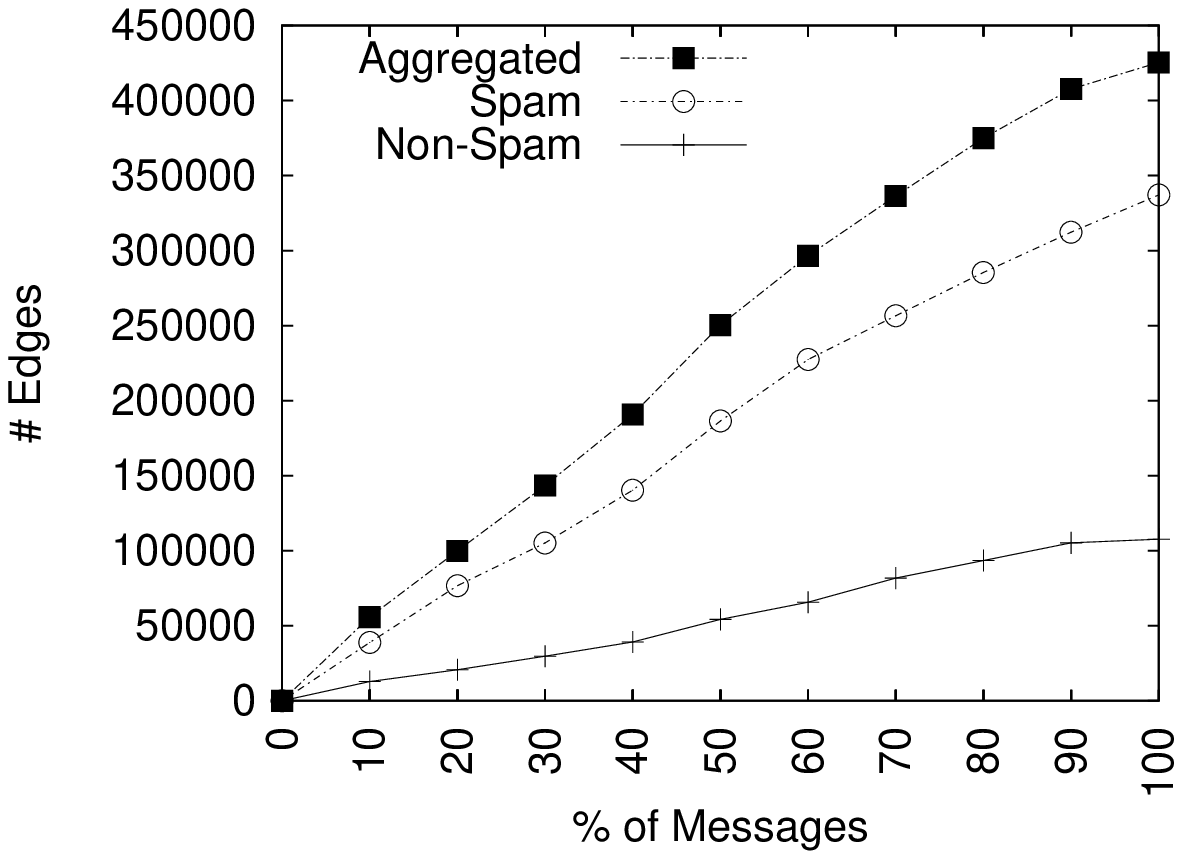} \\      
      (a) User graph&    (b) Domain graph \\    
    \end{tabular}  
  \end{center}  
  \caption{Graph evolution by percentage of messages.}  
  \label{fig:network_evolution}

\end{figure}

A large amount of effort has been devoted recently to creating realistic growth
models for complex networks.  One of the key characteristics of such models is
the evolution of the number of nodes and edges, as well as the
probabilistic connection rules for the new nodes to those already in the graph.
Figure~\ref{fig:network_evolution} shows the evolution of the graph in terms of
number of nodes and edges. %and edge density per node. 
We plot these quantities against percentage of messages 
evaluated for each graph, to avoid the influence of the rate of message arrival,  
which varies with time depending on the type of the traffic being considered
(e.g. the bell shaped behavior for the non spam traffic against the almost constant
rate for spam traffic~\cite{gomes, calzarossa, spam}).   

The growth of the aggregated graph (a composition of the spam and the non-spam graphs) 
results from the growth in both the spam and non-spam
components. The spam subgraph is a much more rapidly growing structure.%, as
%expected from  its larger volume of mail and addresses, recall
%Table~\ref{tab_summary}. 
Over the time of the log we find no saturation effect in these
numbers. Instead the number of addresses and edges grows almost 
linearly with the number of emails. 
%The linear rate of growth is substantially larger for the spam component. 
An eventual saturation in the non-spam component might be expected for longer times. 

%Apart from the fact that the networks grow at two different rates, we found 
%that the node edge density tends to piece wise constants or plateaus 
%during the evolution of the graph.  The first plateau happens in the final part
%of November. The second one, higher than the first, happens in the initial and
%final parts of December. This is due to a  large number of messages related to
%Christmas and new year. The phenomenon is much more apparent in the spam
%traffic and this is due to the commercial aspect of that traffic. 

\begin{figure}[ht]
\vspace{0.25cm}
\footnotesize
  \begin{center}
    {\includegraphics[width=150pt]{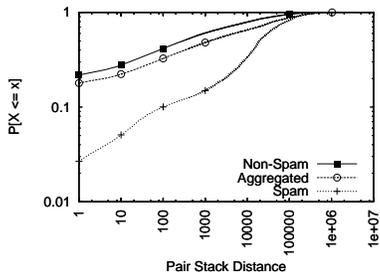}}\\[0.1cm]
    (a) User Graph \\[0.2cm]
    {\includegraphics[width=150pt]{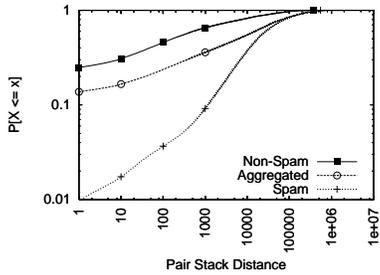}} \\[0.1cm]
    (b) Domain Graph 
  \end{center}
    \caption{Distribution of stack distances for the pairs in the different
      traffics.}
    \label{fig:stack_pairs}
\vspace{0.25cm}
\end{figure}

Another important dynamical graph characteristic is how the weights of edges
evolve, i.e. how the flow of information between nodes varies over time.  
An interesting metric that can be used to measure this is the {\it 
stack distance}~\cite{almeida96characterizing} of connected pairs in terms of the emails they exchange over time. The stack distance measures the number of distinct references between two consecutive
instances of the same object in a stream. We take the total
email log as the stream and each pair sender/receiver as the
object. Ordering of the sender/receiver is disregarded.
Figure~\ref{fig:stack_pairs} shows the pairs' stack distance distributions.
%  for the d2ifferent graphs. 
We see that temporal locality is much stronger in non-spam traffic. This means in practice that legitimate users exchange emails over small concentrations of time.
% As expected, the aggregated traffic tends to approximate the non-spam traffic.  

% retirei essa frase por que como nao consideramos direcionamento a relacao eh
% fraca. Barra
%This diagnostic complements  the information presented in 
%Figure~\ref{fig:reciprocity}  on communication reciprocity, that considered the
%direction of the communications. 

We were also interested in studying how do the nodes communicate with their
peers in terms of the number of messages.  Because of the impersonal nature of
spam we expect that spam senders communicate in a more structured way with
their recipients. 
% while legitimate people communicate with greater variability. 
Not only will legitimate senders show more variation in  the number of messages
they send to each person in their out sets, they will also show variability of
the messages themselves  in terms of their sizes. In order to quantify these
effects we evaluated the normalized entropy of the in and out flows for  each
node, defined as  
\begin{equation}
  H(x) = \frac{\sum_{y \in OS(x) }{-p(y) * log(p(y))}}{log(|S(x)|)},
\end{equation}
where $p(y)$ is the probability of $y$ receiving a message from $x$ and  and $|S(x)|$
is the number of unique elements in the set being considered.

\begin{figure}[ht]
\vspace{0.30cm}
\footnotesize
    \begin{center}
      {\includegraphics[width=150pt]{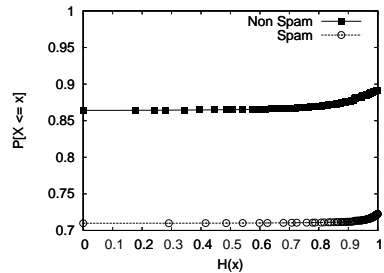}} \\[0.1cm]
      (a) User Graph \\[0.2cm]
      {\includegraphics[width=150pt]{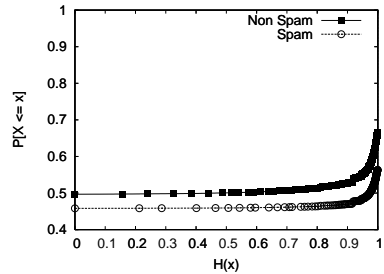}}\\[0.1cm]
      (b) Domain Graph \\[0.2cm]
   \end{center}
    \caption{Distribution of entropy of the number of messages in the flow of
      e-mails for the aggregated graph.}
    \label{fig:flow_emails}

%%%%\vspace{-0.15cm}
\end{figure}

Figure~\ref{fig:flow_emails} shows the normalized entropy for the
out flow of the nodes in the different sender classes for the
aggregated graphs. As expected, spammers
communicate with their recipients with much less variability (higher entropy). 
%The outgoing flow of the nodes that send both non-spam and spam
%%presents much more variability than the others. This happens since the lists of the
%people that used the forged email addresses to send spams is different than the
%list of people to which the real user sends messages. 
A similar analysis was conducted considering the bytes that each node
sends with similar results.  

%We begin by analyzing the amount of information exchanged among users or flow,  
%%In order to study flow between pairs of nodes and between a node and its incoming or outgoing set, we use 
%in the labeled graphs defined in section \ref{graphs}.
%Figures \ref{fig:flow_pairs} a-b show the distribution of flow measured in number of messages exchanged for user and domain graphs. We see that in both graphs, pairs in spam traffic exchange less messages than in non spam traffic. Similar results were found when considering graphs labeled with number of bytes. This behavior is consistent with the results of other metrics 
%discussed above that suggest that a spam sender distributes his load over many addresses, resulting in a sm,aller number of messages per sender id.
%This result joint with smaller size of spam emails showed in \cite{gomes}, explain the less intense flow, too, in number of bytes between pairs of nodes in both graphs.

\section{Related Work}
\label{sec:related-work}

Several studies have recently analyzed the statistical properties of email workloads
~\cite{spam,spamimc04,gomes,calzarossa,priority}. These studies consider the
messages as a flow and study metrics such as inter-arrival times, e-mail sizes, 
and number of recipients per e-mail. Although spam and legitimate email show
differences in terms of these metrics little has been done about using them to filter out 
spam. The work of the present manuscript takes a different tack by creating a
graph theoretical higher level representation of email traffic and attempting
to differentiate spam from legitimate email in this abstraction. 
We believe that this approach, based on graph theoretical metrics,
proves to be much better suited to the filtering problem. 

Other recent papers have focused on models of email traffic as
graphs~\cite{emailnetcombat, spammachines}. For example in
Ref. ~\cite{emailnetcombat} a graph is created representing the email traffic
captured by the mailbox of an individual user. The subsequent structural
analysis is based on the fact that such a network possesses several
disconnected components. The clustering coefficient of each of  these
components is then used to characterize messages as spam or non-spam. Their
results show that 53\% of the messages were classified using the proposed
approach and they obtained 100\% of accuracy in this subset. Our graphs are based
 on a different type of dataset, i.e. the logs of SMTP servers, and as such do
not take the perspective of the individual user.  As a result for our data set
the approach proposed in~\cite{emailnetcombat} can not be used successfully
since there is a giant SCC in all of the graphs shown. In~\cite{spammachines}
the authors used the approach of detecting machines that behave as spam senders
by analyzing a border flow graph of sender and recipient machines. Moreover,
they analyzed the evolving graph structures over a period of time, based on a
single metric using the HITS algorithm. Our workload differs from theirs since
we do not have access to the underlying overlay network formed by email relays.

\section{Conclusions}
\label{sec:concl-future-works}

In this paper, we have shown that legitimate and spam email graphs differ in 
two fundamental classes of characteristics: structural, which capture the graphs' architecture, and dynamical, concerning node communication and graph evolution. 

Structurally we showed that spam and non spam subgraphs are characterized by
different distributions of the clustering coefficient of their nodes.
Legitimate email users display on average higher clustering coefficients than
spam senders.  Node visitation probability is a measure of the centrality of 
a node relative to other nodes in the graph. 
%The higher the node visitation the higher the node betweeness.  
Legitimate email nodes have higher visitation probability than spam nodes.  We
also defined a new metric called communication reciprocity. It measures the
probability that a node receives a response from any of its addressees.  
There is a strong difference in the probability distributions 
of the communication reciprocity in the legitimate and spam graphs;  
legitimate nodes have a much higher probability of being responded to. 
Another metric introduced in this paper is the email asymmetry set, which represents the difference between the sets of in and out edges of a node. We showed that there is a strong correlation between the size of asymmetry sets and the number of spammers in the set. 
%The dynamical properties concerns the evolution of nodes and edges of a growing graph.  
Dynamically the spam graph grows much faster than the legitimate email
graph. The legitimate email graph grows more slowly both in the number of nodes
and edges, manifesting the higher stability of relations in a social group. Two
other dynamical metrics, entropy and stack distance, are used to reveal the
temporal characteristics of communication among nodes. Spam nodes display a
much higher entropy than legitimate email users, and a much longer stack
distance. 

We have shown that differences in both classes of graph characteristics can be explained by the same hypothesis, namely that legitimate email graphs reflect real social networks, while spam graphs are technological networks, devoid of a sense of community. Although no single metric can unequivocally differentiate legitimate emails from spam, the combination of several graphical measures paint a clear picture of the 
%discussed in the paper contributes to increase the understanding of  the 
processes whereby legitimate and spam email are created.  For this reason they can be used to augment the  effectiveness of mechanisms to detect illegitimate emails.

\subsubsection*{Acknowledgements}
Luiz H. Gomes is supported by Banco Central do Brasil.

%\bibliographystyle{abbrv}
%\bibliography{../bib-file/spam}

\footnotesize
 
\begin{thebibliography}{10}

\bibitem{almeida96characterizing}
V.~Almeida, A.~Bestavros, M.~Crovella, and A.~de~Oliveira.
\newblock Characterizing reference locality in the {WWW}.
\newblock In {\em Proc. of the {IEEE} Conference on Parallel and Distributed
  Information Systems ({PDIS})}, Miami Beach, FL, 1996.

\bibitem{technologicalnetworks}
J.~Balthrop, S.~Forrest, M.~E.~J. Newman, and M.~M. Williamson.
\newblock Technological networks and the spread of computer viruses.
\newblock In {\em Computer Science}, April 2004.

\bibitem{calzarossa}
L.~Bertolotti and M.~C. Calzarossa.
\newblock Workload characterization of mail servers.
\newblock {\em Proc. of SPECTS 2000, Vancouver, Canada}, July 2000.

\bibitem{emailnetcombat}
P.~O. Boykin and V.~Roychowdhury.
\newblock Personal email networks: An effective anti-spam tool.
\newblock http://www.arxiv.org/abs/cond-mat/0402143, February 2004.

\bibitem{pagerank}
S.~Brin and L.~Page.
\newblock The anatomy of a large-scale hypertextual web search engine.
\newblock In {\em Proc. of the 9th International World wide Web Conference},
  1998.

\bibitem{spam}
L.~F. Cranor and B.~A. LaMacchia.
\newblock Spam!
\newblock In {\em Communications of the ACM}, 1998.

\bibitem{spammachines}
P.~Desikan and J.~Srivastava.
\newblock Analyzing network traffic to detect e-mail spamming machines.
\newblock Technical Report 180, Army High Performance Computing Research Center
  TECHNICAL REPORT, 2004.

\bibitem{evolutionnet}
S.~N. Dorogvtsev and J.~F.~F. Mendes.
\newblock Evolution of networks.
\newblock In {\em Advances in Physics}, pages 1079--1187, 2002.

\bibitem{gomes}
L.~H. Gomes, C.~Cazita, J.~Almeida, V.~A.~F. Almeida, and W.~M. Jr.
\newblock Characterizing a spam traffic.
\newblock In {\em Proc. of the 4th ACM SIGCOMM conference on Internet
  measurement}, 2004.

\bibitem{NYTarticle}
S.~Hansell.
\newblock Dozens charged in push against spam and scams.
\newblock http://www.nytimes.com/2004/08/25/, August 2004.

\bibitem{spamimc04}
J.~Jung.
\newblock An empirical study of spam traffic and the use of dns black lists.
\newblock In {\em Proc. of the 4th ACM SIGCOMM conference on Internet
  measurement}, 2004.

\bibitem{messageLabs}
M.~Labs.
\newblock Message labs home page.
\newblock http://www.messagelabs.co.uk/, 2005.

\bibitem{emailnetwork}
M.~E. Newman, S.~F. S., and J.~Balthrop.
\newblock Email networks and the spread of computer viruses.
\newblock In {\em Physical Review E}, September 2002.

\bibitem{socialxother}
M.~E.~J. Newman and J.~Park.
\newblock Why social networks are different from other types of networks.
\newblock In {\em Physical Review E (Statistical, Nonlinear, and Soft Matter
  Physics)}, September 2003.

\bibitem{spamassassin}
Spamassassin.
\newblock http://www.spamassassin.org, 2004.

\bibitem{spamhaus}
Spam haus project.
\newblock http://www.spamhaus.org, 2005.

\bibitem{priority}
R.~D. Twining, M.~M. Willianson, M.~Mowbray, and M.~Rahmouni.
\newblock Email prioritization: Reducing delays on legitimate mail caused by
  junk mail.
\newblock In {\em Proc. Usenix Annual Technical Conference}, Boston, MA, June
  2004.

\end{thebibliography}

\end{document}